# Can the observer know the state of Schrödinger's cat without opening the box?

Lizhi Xin[1] and Houwen Xin[2,*]

[1]Building 59, 96 Jinzhai Road, Hefei, Anhui, P. R. China
[2]Department of Chemical physics USTC, Hefei, Anhui, P. R. China
[*]hxin@ustc.edu.cn

**ABSTRACT**

In order to know if the Schrödinger's cat is alive or dead without opening the box, observers have to play a game with nature. The observers have to "guess" (with degrees of belief) the state of the cat due to incomplete information; in other words, the observers' decision has to be made under uncertainty. We propose a quantum expected value theory for decision-making under uncertainty. Value operator is proposed to guide observers to choose corresponding actions based on their subjective beliefs through objective quantum world by maximizing the value from the measured historical results. The value operator, as a quantum decision tree, can be constructed from both quantum gates and logic operations. Genetic programming is applied to optimize quantum decision trees.

## Introduction

In 1935, Erwin Schrödinger created his famous thought experiment involving a cat that is simultaneously both dead and alive to show how absurd quantum physics is.

"A cat is penned up in a steel chamber, along with the following device (which must be secured against direct interference by the cat): in a Geiger counter there is a tiny bit of radioactive substance, so small, that perhaps in the course of the hour one of the atoms decays, but also, with equal probability, perhaps none; if it happens, the counter tube discharges and through a relay releases a hammer which shatters a small flask of hydrocyanic acid. If one has left this entire system to itself for an hour, one would say that the cat still lives if meanwhile no atom has decayed. The psi-function of the entire system would express this by having in it the living and dead cat (pardon the expression) mixed or smeared out in equal parts."

By entangling an atom from the microscopic world with a cat from the macroscopic world, Schrödinger asked a question that is difficult to answer with the Copenhagen Interpretation of quantum mechanics: where is the clear boundary between the quantum world and the classical world? It was this cat that multiple quantum interpretations[1-7] was proposed and the debate regarding it still goes on. Schrödinger continued to state in the same paper:

"We saw that the indeterminacy is not even an actual blurring, for there are always cases where an easily executed observation provides the missing knowledge. So what is left? From this very hard dilemma the reigning doctrine rescues itself by having recourse to epistemology. We are told that no distinction is to be made between the state of a natural object and what I know about it, or perhaps better, what I can know about it if I go to some trouble. Actually - so they say - there is intrinsically only awareness, observation, measurement."

Now if we ask this question: is it possible that the observer can "know" the state of cat (alive or dead) without opening the box? In other words, is it possible that the observer can have a reasonable expectation of the cat's state by learning historical results of repeated measurements on copies of the same system (atom + cat)?

Whether atoms decayed or not is unpredictable, and it is the indeterminacy of objective world (atom) that causes the uncertainty of people's subjective beliefs (guessing if the cat is alive or dead). For observers to believe if the cat is alive or dead is actually a problem of decision-making under uncertainty due to incomplete information. We propose a quantum expected value decision theory[8] to deal with observers' decision-making problem[9-11], and quantum genetic programming is applied to evolve "satisfactory" strategies for observers to "know" (with degrees of belief) the state of the cat as best possible based on quantum expected value.

## Quantum expected value (qEV)

| State \ Action | $\varphi_1$ | $\varphi_2$ |
|---|---|---|
| $a_1$ | $r_{11}$ | $r_{12}$ |
| $a_2$ | $r_{21}$ | $r_{22}$ |

Table 1 State-action-value decision table

The observer subjectively chooses an action $a_i \in \{a_1, a_2\}$ where the atom's objective state is in $\varphi_j \in \{\varphi_1, \varphi_2\}$ when a decision is made, and the result (matrix $r_{ij}$) depends on both the state of the atom and choice of observer's brain shown in table 1. The state of the atom describes the objective world; it can be represented by the superposition of all possible states in terms of the Hilbert state space[12-14] as in (1). The observer's mental state describes the subjective world; we postulate that when the observer is undecided in making a decision, it can be represented by superposition of all possible actions as in (2). The information of the quantum world is incomplete; the result of the observer's decision is uncertain and it can be represented by a mixed state's density operator as a value operator in (3). Value operator is a sum of projection operators which projects the observer's degrees of belief onto an action of choice. Quantum expected value can be represented as in (4).

$$|\psi\rangle_{atom} = c_1|\varphi_1\rangle + c_2|\varphi_2\rangle \quad |c_1|^2 + |c_2|^2 = 1 \tag{1}$$

$$|\phi\rangle_{mental} = \mu_1|a_1\rangle + \mu_2|a_2\rangle \quad |\mu_1|^2 + |\mu_2|^2 = 1 \tag{2}$$

$$\hat{V} = p_1|a_1\rangle\langle a_1| + p_2|a_2\rangle\langle a_2| \quad p_1 + p_2 = 1 \tag{3}$$

$$qEV = \langle\psi|\hat{V}|\psi\rangle = (c_1^*\langle\varphi_1| + c_2^*\langle\varphi_2|)(p_1|a_1\rangle\langle a_1| + p_2|a_2\rangle\langle a_2|)(c_1|\varphi_1\rangle + c_2|\varphi_2\rangle)$$
$$= p_1\omega_1|\langle a_1|\varphi_1\rangle|^2 + p_1\omega_2|\langle a_1|\varphi_2\rangle|^2 + p_2\omega_1|\langle a_2|\varphi_1\rangle|^2 + p_2\omega_2|\langle a_2|\varphi_2\rangle|^2$$
$$= p_1\omega_1 r_{11} + p_1\omega_2 r_{12} + p_2\omega_1 r_{21} + p_2\omega_2 r_{22}$$
$$= \sum_{i=1,2} p_i \sum_{j=1,2} \omega_j r_{ij} \tag{4}$$

Where $|\varphi_1\rangle$ indicates a state in which the atom has not decayed and $|\varphi_2\rangle$ indicates a state in which the atom has decayed; $|a_1\rangle$ represents the observer's action to believe the cat is alive and $|a_2\rangle$ represents the observer's action to believe the cat is dead; $p_i = |\mu_i|^2$ is a observer's subjective probability in choosing an action $a_i$, subjective probability represents the observer's degrees of belief in a single event; $\omega_j = |c_j|^2$ is the objective frequency at which state of the atom is in $\varphi_j$, objective frequency represents the statistical results of multiple occurrences of objective states; matrix $r_{ij} = |\langle a_i|\varphi_j\rangle|^2$ is the value when the decision was made, in which the observer choose an action $a_i$ where atom's state is in $\varphi_j$. The different actions that the observers took lead to different value; in other words; the value is "created" based on both observers' subjective beliefs and objective natural states.

The process of decision-making can be interpreted as a game between observers and nature; the final decision is equivalent to a "quantum measurement" performed on the mental state by the observers' brain, which choose an action $a_i$ with probability $p_i$ as in (5) (observer's mental state is transformed from a pure state into a mixed state, and then one of available actions is selected by the brain). Based on the information that the brain has before a decision is made, there is only the possibility of selecting an action with a subjective probability (degrees of belief), and it is the decision-making process that makes the potential possibility a reality. The decision process of an observer can be simulated by the continuous evolution of the value operator according to the environment (information). Information is the essence of people's subjective beliefs just like energy is the essence of the objective world. Valuable information can reduce uncertainty.

$$D: \hat{\rho} = |\phi\rangle\langle\phi| \rightarrow \hat{V} = p_1|a_1\rangle\langle a_1| + p_2|a_2\rangle\langle a_2| \xrightarrow{decision} |a_i\rangle\langle a_i|, i = 1,2 \tag{5}$$

## Quantum decision tree (qDT)

A value operator can be constructed from basic quantum gates[15-16] and logic operations to form a qDT. The qDT composes of different nodes and branches. There are two types of nodes, non-leaf nodes and leaf nodes. The non-leaf nodes are composed of the operation set F as in (6); the leaf nodes are composed of the data set T as in (7)-(8). The construction process of a qDT is to randomly select a logic symbol from the operation set F as the root of the qDT, and then grows corresponding branches according to the nature of the operation symbol and so on until a leaf node is reached.

$$F = \{+(ADD), \quad *(MULTIPLY), \quad //(OR)\} \tag{6}$$

$$T = \{H, X, Y, Z, S, D, T, I\} \tag{7}$$

$$\begin{cases} H = \frac{1}{\sqrt{2}}\begin{bmatrix}1 & 1\\1 & -1\end{bmatrix} \; X = \begin{bmatrix}0 & 1\\1 & 0\end{bmatrix} \; Y = \begin{bmatrix}0 & -i\\i & 0\end{bmatrix} \; Z = \begin{bmatrix}1 & 0\\0 & -1\end{bmatrix} \\ S = \begin{bmatrix}1 & 0\\0 & i\end{bmatrix} \; D = \begin{bmatrix}0 & 1\\-1 & 0\end{bmatrix} \; T = \begin{bmatrix}1 & 0\\0 & e^{i\pi/4}\end{bmatrix} \; I = \begin{bmatrix}1 & 0\\0 & 1\end{bmatrix} \end{cases} \tag{8}$$

The qDT of a value operator is a 2x2 matrix, and the value operator needs to be diagonalized first and then normalized to get probability $p_1$ and $p_2$ as in (9)-(10).

$$\hat{V} = \begin{bmatrix}V_{11} & V_{12}\\V_{21} & V_{22}\end{bmatrix} \xrightarrow{\text{diagonalization}} \begin{bmatrix}\lambda_1 & 0\\0 & \lambda_2\end{bmatrix} \xrightarrow{\text{normalization}} \begin{bmatrix}p_1 & 0\\0 & p_2\end{bmatrix} = p_1|a_1\rangle\langle a_1| + p_2|a_2\rangle\langle a_2| \tag{9}$$

$$|a_1\rangle = \begin{bmatrix}1\\0\end{bmatrix}, |a_2\rangle = \begin{bmatrix}0\\1\end{bmatrix}; \; |a_1\rangle\langle a_1| = \begin{bmatrix}1 & 0\\0 & 0\end{bmatrix}, |a_2\rangle\langle a_2| = \begin{bmatrix}0 & 0\\0 & 1\end{bmatrix} \tag{10}$$

## Quantum genetic programming (qGP)

A tree structure is used for encoding by genetic programming (GP)[17-20]; the qDT can be optimized by the qGP. The purpose of qGP iterative evolution is to find a satisfactory qDT through learning historical data. The learning rule is as follows:

a) If the cat is alive ($\varphi_1$) and an observer believes the cat is alive ($a_1$), the value $V_k$ is positive and the observer would get an award (v); if an observer believes the cat is dead ($a_2$), the value $V_k$ is negative, observer would get a penalty (-v).

b) If the cat is dead ($\varphi_2$) and an observer believes the cat is dead ($a_2$), the value $V_k$ is positive, the observer would get an award (v); if an observer believes the cat is alive ($a_1$), the value $V_k$ is negative, observer would get a penalty (-v).

An optimization problem mainly includes the selection of evaluation function and the acquisition of optimal solution. The evaluation function of qDT is a fitness function $f_{\text{fitness}}$ (14) based on observed value $V_k$ (13), and the optimal solution is obtained through continuous evolution by using selection, crossover, mutation as in (15) and implemented by qGP algorithm.

$$<V_k> = p_i\omega_j\langle\varphi_j|A_i|\varphi_j\rangle, \quad A_i = |a_i\rangle\langle a_i| \text{ is Von Neumann's projection operator} \tag{11}$$

$$\langle\varphi_j|A_i|\varphi_j\rangle = \langle\varphi_j||a_i\rangle\langle a_i||\varphi_j\rangle = |\langle a_i|\varphi_j\rangle|^2 = r_{ij} = \begin{cases}v, & i = j\\-v, & i \neq j\end{cases} \tag{12}$$

For the kth measurement, the quantum expected value is:

$$<V_k> = \begin{cases} p_1\omega_1|\langle a_1|\varphi_1\rangle|^2 = p_1\omega_1 r_{11} = p_1\omega_1 v, & |\phi\rangle_{\text{mental}} = |a_1\rangle \text{ and } |\psi\rangle_{\text{atom}} = |\varphi_1\rangle\\ p_1\omega_2|\langle a_1|\varphi_2\rangle|^2 = p_1\omega_2 r_{12} = -p_1\omega_2 v, & |\phi\rangle_{\text{mental}} = |a_1\rangle \text{ and } |\psi\rangle_{\text{atom}} = |\varphi_2\rangle\\ p_2\omega_1|\langle a_2|\varphi_1\rangle|^2 = p_2\omega_1 r_{21} = -p_2\omega_1 v, & |\phi\rangle_{\text{mental}} = |a_2\rangle \text{ and } |\psi\rangle_{\text{atom}} = |\varphi_1\rangle\\ p_2\omega_2|\langle a_2|\varphi_2\rangle|^2 = p_2\omega_2 r_{22} = p_2\omega_2 v, & |\phi\rangle_{\text{mental}} = |a_2\rangle \text{ and } |\psi\rangle_{\text{atom}} = |\varphi_2\rangle \end{cases} \tag{13}$$

$$f_{\text{fitness}} = \sum_{k=0}^{N} <V_k> \tag{14}$$

$$\text{qDT} \xrightarrow{\text{evolution}} \underset{\text{qDT} \in (F \cup T)}{\text{argmax}} (f_{\text{fitness}}) \tag{15}$$

**qGP algorithm**

*Input*:
- Historical data set $\{d_k = (\varphi_k, v_k), k = 0, \cdots, N\}$ (each sample consists of an atom's state and value for observers).
- Setting
1) Operation set $F = \{+, *, //\}$
2) Data set $T = \{H, X, Y, Z, S, D, T, I\}$, eight basic quantum gates
3) Crossover probability = 60~90%; Mutation probability = 1~10%.

*Initialization*:
- Population: randomly create 100 ~ 500 qDTs.

*Evolution*:
- for $i = 0$ to $n$, $n = 50~100$ generations
a) Calculate fitness for each qDT based on historical data set.
b) According to the quality of fitness:
    i. Selection: selecting parent qDTs.
    ii. Crossover: generate a new offspring using the roulette algorithm based on crossover probability.
    iii. Mutation: randomly modify parent qDT based on mutation probability.

*Output*:
- A qDT of the best fitness.

## Results

The historical measurement results of Schrödinger's cat thought experiment are simulated by a computer program which randomly generates N (=10000) results (atom decayed or not decayed). We can define a random variable x to represent the fluctuation of the cat's state as in (16). The generated data series is $\{x_1, \cdots, x_k, x_{k+1}, \cdots, x_N\}$; if the atom has not decayed then variable x increases by 1, else variable x decreases by 1. The uncertainty of the cat's state is represented by the volatility of variable x as shown in Figure 1. The value obtained by the observer after each action is as in (17).

$$x_{k=0} = 0; \quad x_{k+1} = x_k + \begin{cases} -1, & \text{if decayed} \\ 1, & \text{if not decayed} \end{cases} \quad k = 1, \cdots, N \tag{16}$$

$$v_k = |x_{k+1} - x_k| \tag{17}$$

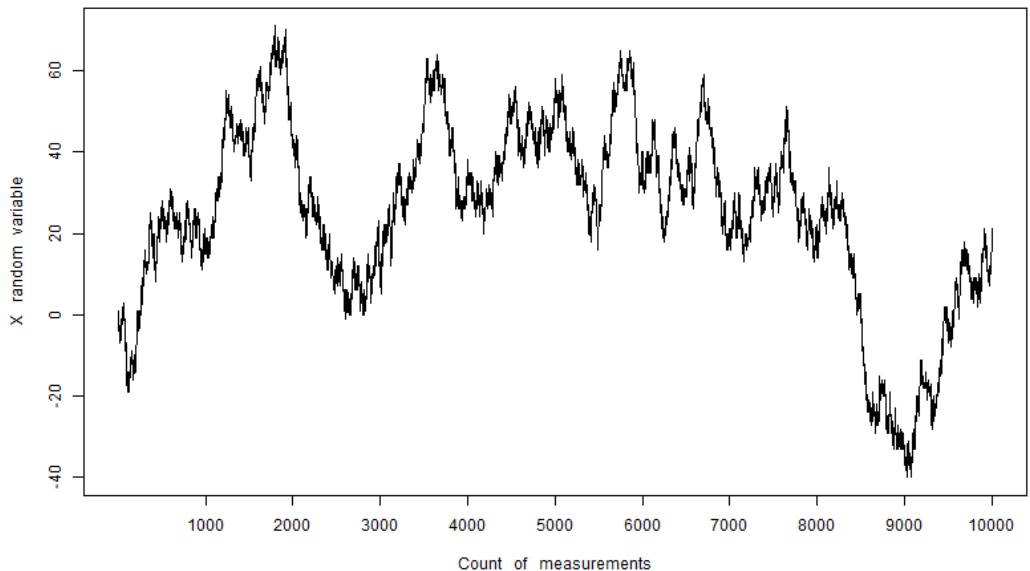

**Figure 1 Fluctuation of states of the cat.**

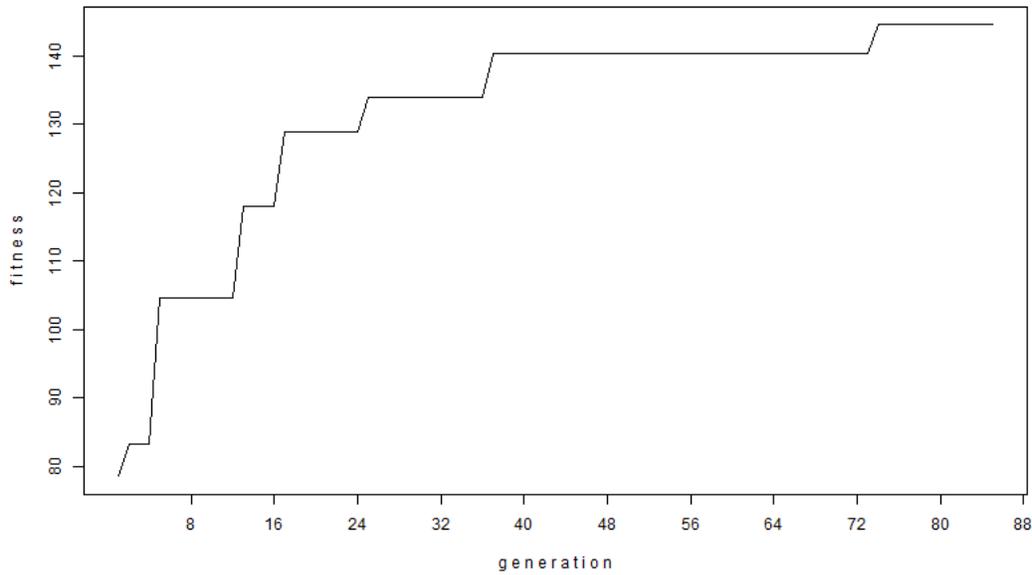

**Figure 2 Evolution of the qDT**

Through learning the historical data set, an optimized qDT is obtained by qGP. The evolution of the qDT is shown in Figure 2 and the optimized qDT after 88 generations of evolution is shown in Figure 3. Based on the qDT as in (18) (Figure 3), there are four strategies with different subjective degrees of belief that the observer can take. At any given moment the observer's degrees of belief are unknown, the qDT which simulates observer's degrees of belief can be interpreted as a mixed strategy with four different strategies $\{S_1, S_2, S_3, S_4\}$ for the observer, and the final decision is made by "quantum projection measurement" which the observer's brain selects an action ($a_1$: believe cat is alive or $a_2$: believe cat is dead) with degrees of belief from one of four available strategies $S_i = \{S_1, S_2, S_3, S_4\}$.

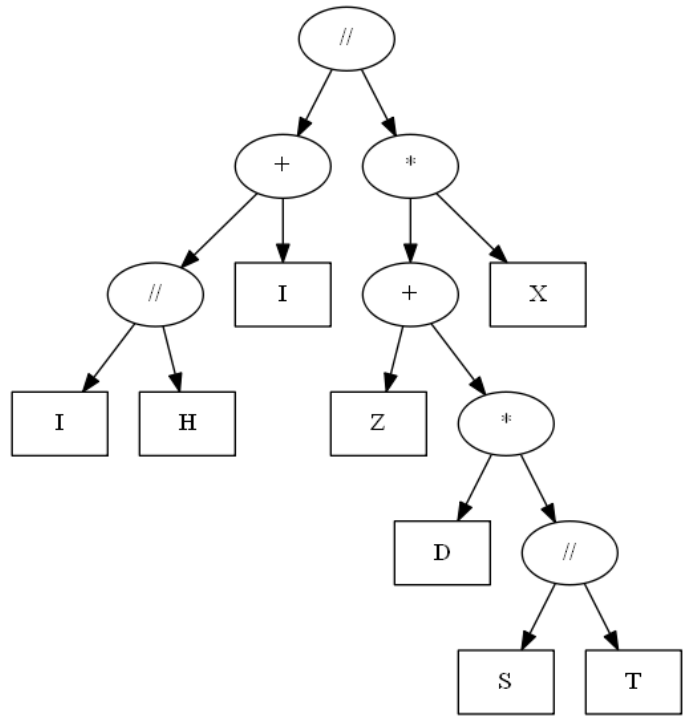

**Figure 3 An optimized qDT.**

$$\text{qDT} = \Big(\big((I//H) + I\big) // \big((Z + (D * (S//T))) * X\big)\Big) \tag{18}$$

- $S_1 = (I + I) \to \hat{V} = 0.5|a_1\rangle\langle a_1| + 0.5|a_2\rangle\langle a_2|$ (50% belief cat is alive, 50% belief cat is dead)
- $S_2 = (H + I) \to \hat{V} = |a_1\rangle\langle a_1|$ (100% belief cat is alive)
- $S_3 = \big((Z + (D * S)) * X\big) \to \hat{V} = 0.05|a_1\rangle\langle a_1| + 0.95|a_2\rangle\langle a_2|$ (5% belief cat is alive, 95% belief cat is dead)
- $S_4 = \big((Z + (D * T)) * X\big) \to \hat{V} = 0.17|a_1\rangle\langle a_1| + 0.83|a_2\rangle\langle a_2|$ (17% belief cat is alive, 83% belief cat is dead)

The subjective degrees of belief of the first 100 actions the observer took are shown in Figure 4. Detailed information of the first ten actions the observer took is shown in Table 2. For the first action, strategy $S_3$ was applied by the observer who believes that the cat is dead with 95% degrees of belief, the observer got it wrong because the cat is alive when the box is opened; for the fourth action, strategy $S_3$ was applied by the observer who believes that the cat is dead with 95% degrees of belief, this time the observer got it right because the cat is dead when the box is opened; for the eighth action, strategy $S_1$ was applied by the observer who believe that the cat is alive with 50% degrees of belief, the observer got it wrong because the cat is dead when the box is opened; for the tenth action, strategy $S_2$ was applied by the observer who believe that the cat is alive with 100% degrees of belief, the observer got it right this time because the cat is alive when the box is opened. By using this particular qDT, the rate of success for the observer is 52%. If $N \to \infty$, rate of success $\to$ 50% (if the atom decays or not is completely random).

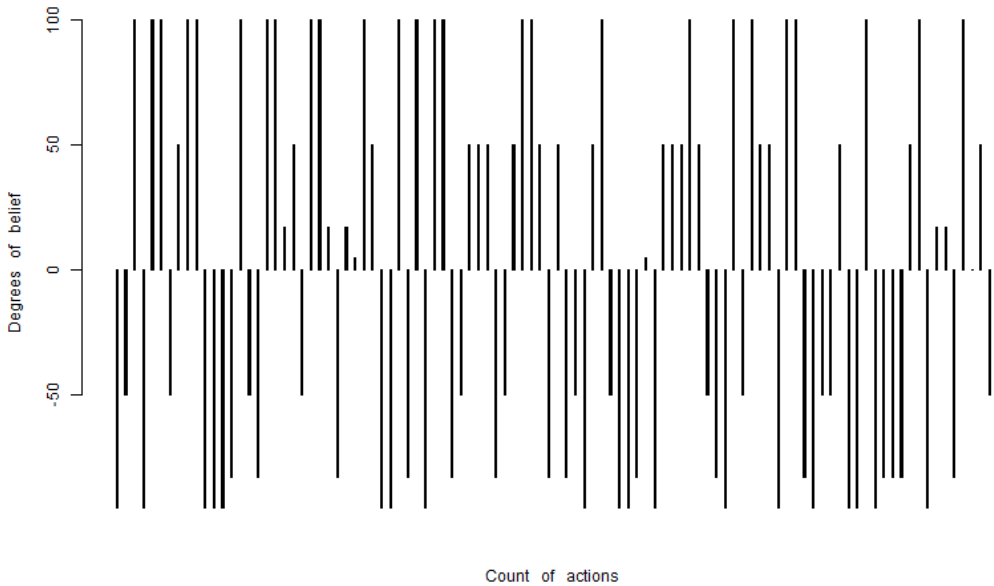

**Figure 4 A Observer's degrees of belief of first 100 actions (positive: believe cat is alive, negative: believe cat is dead).**

| Cat's state | Observer's action selected | Degrees of belief | Strategy | Value |
|---|---|---|---|---|
| $\varphi_1$: alive | $a_2$: believe cat is dead | 95% | $S_3$ | -1 |
| $\varphi_2$: dead | $a_2$: believe cat is dead | 50% | $S_1$ | 1 |
| $\varphi_2$: dead | $a_1$: believe cat is alive | 100% | $S_2$ | -1 |
| $\varphi_2$: dead | $a_2$: believe cat is dead | 95% | $S_3$ | 1 |
| $\varphi_2$: dead | $a_1$: believe cat is alive | 100% | $S_2$ | -1 |
| $\varphi_1$: alive | $a_1$: believe cat is alive | 100% | $S_2$ | 1 |
| $\varphi_2$: dead | $a_2$: believe cat is dead | 50% | $S_1$ | 1 |
| $\varphi_2$: dead | $a_1$: believe cat is alive | 50% | $S_1$ | -1 |
| $\varphi_1$: alive | $a_1$: believe cat is alive | 100% | $S_2$ | 1 |
| $\varphi_1$: alive | $a_1$: believe cat is alive | 100% | $S_2$ | 1 |

**Table 2 The first ten actions.**

# Discussion

Measuring instruments record a large amount of data through interaction with the quantum world. It is through the study of the recorded historical data that observers gradually understand the objective quantum world and build a simplified subjective "world model" in the brain. Observers make decisions by considering both the quantum world's objectivity and the subjectivity of their beliefs. The observers' subjective mental state transitions from a pure state to a mixed state to complete a decision-making as in (5), and the observed value of the decision is obtained. Observed value is a bridge between objective quantum world and observers' subjective beliefs. By maximizing the observed value, the decision model can be continuously improved by quantum genetic programming which obtains a satisfactory set of decision strategies for observers.

A living creature will certainly not be in a superposition of dead and alive, nor will it collapse because of the last glimpse of the observer, or two different worlds existing at the same time, one for a living cat and another one for a dead cat. For observers, Schrödinger's cat paradox is caused by uncertainty in decision-making due to incomplete information. It may never be possible to accurately predict whether a cat will be alive at the next moment. There are infinite "value" paths pointing to the end, we may never find the best "value" path, and all we can do is find a satisfactory "value" path as best possible through evolution.

For a large amount of repeated measurements, there is an objective frequency (50% the cat is alive or dead); but for a single measurement there is no so-called objective probability, the cat is either alive or dead; and in this case, only the observer has subjective degrees of belief that the cat is alive or dead before the box is opened.

The fundamental reason for the unpredictability of the states of the cat for observers is due to incomplete information regarding the quantum world, and there is a quantum interpretation problem[21-28], that is, what is the essence of indeterminacy in the quantum world?

As Schrödinger said：

"It is typical of these cases that an indeterminacy originally restricted to the atomic domain becomes transformed into macroscopic indeterminacy, which can then be resolved by direct observation. That prevents us from so naively accepting as valid a 'blurred model' for representing reality. In itself it would not embody anything unclear or contradictory. There is a difference between a shaky or out-of-focus photograph and a snapshot of clouds and fog banks".